\documentclass[twocolumn,showpacs,prl,amsmath,amssymb]{revtex4}
\usepackage{epsfig}
\begin{document}

\title{The double-slit and the EPR experiments: A paradox-free kinematic description}
\author{A. \surname{Kryukov}} 
\affiliation{Department of Mathematics, University of Wisconsin Colleges, 780 Regent Street, Madison, WI 53708} 
\date{\today}
\begin{abstract}
The paradoxes of the double-slit and the EPR experiments with particles are shown to originate in the implicit assumption that the particles are always located in the classical space. It is demonstrated that there exists a natural substitute for this assumption that provides a method of resolving the paradoxes.
\end{abstract}


\maketitle

\section{The double-slit experiment}
\setcounter{equation}{0}

Two physical experiments have captured the paradoxical nature of quantum mechanics in an elementary yet essentially complete way: the double-slit experiment and the EPR experiment.
Few experiments in the history of science have generated so many ferocious debates, prompted so many controversial interpretations and at the end left us with such a deep feeling of discomfort with the current state of affairs. 
Recall the mysterious notions of the ``wave-particle duality'' and of ``quantum non-locality'' originating in the experiments. 

The double-slit and the EPR experiments are essentially similar. In fact, in both of them one deals with superpositions of the classically meaningful states and those superpositions are the source of all controversies in the theory. 
Accordingly, there is essentially only one mystery in quantum mechanics: the existence of superpositions of classically meaningful states. So, to understand quantum mechanics is to understand superpositions of states. 

In particular, the electron in the double-slit experiment is in a superposition of states that describe electron passing through one of the slits. There are three logical possibilities for such an electron:
\begin{description}
\item [(A)]
the electron passes through both slits
\item [(B)]
the electron passes through only one of the slits
\item [(C)]
the electron does not pass through the slits at all
\end{description}
Which of these possibilities is realized? 

The most common answer within the quantum community is (A).
Namely, one says that the electron in the given state behaves like a wave, rather than a particle. The wave passes through both slits at once causing an interference pattern behind the slits. 
One goes on to say that the wave function gives a complete quantum description of the electron's state. It yields the probability of finding the electron near an arbitrary spatial point (Born's rule). 
It is normally assumed that the electron itself is real (physical), while the wave function is not. 
The sketched position is due primarily to Bohr and it serves a basis for the famous Copenhagen interpretation.

In this approach the electron in the given state has no definite position. In particular, it cannot be near a single slit as otherwise the wave function would be concentrated at that slit. Paradoxically, the fact that the wave function vanishes away from the slits seems to indicate that the electron must be near the pair of the slits. It follows that the electron splits somehow into two parts. 
The density of the resulting ``electron cloud'' coincides with the square of the modulus of the wave function. This relationship of the wave function with the physically meaningful density contradicts its earlier mentioned non-physicality. 

However, whenever necessary, the Copenhagen interpretation distances itself from such problematic conclusions and logical contradictions. 
Instead, it retreats to the view that one should only be concerned with measurements, in which no electron parts can ever be observed and no need for a physical wave function ever arises. This runaway argument may indeed eliminate the problem, but it leaves one with a feeling of guilt for the ostrich-like behavior.
 
Einstein on the other hand maintained that the possibility (B) is realized. That is, the electron in the experiment goes through only one of the slits, but the standard quantum mechanics does not tell us the whole story. His famous question``Do you really think the moon isn't there if you aren't looking at it?'' pushed for the development of the more detailed, ``hidden variables'' theories. However, as well known after Bell, those theories can only be reconciled with experiment if they admit some form of 
``action at a distance'' i.e., nonlocality.
Ironically, one of the main motivations for Einstein to promote that type of approach was to eliminate nonlocality from the theory. 

Following Bohr and Einstein, the possibilities (A) and (B) were extensively explored and various more 
advanced interpretations of these possibilities were considered. 
On the other hand, the possibility (C) has not been seriously investigated. 
If realized, this possibility would mean that
the electron disappears somehow between the source and the screen with the slits and then reappears on the other side of the screen, when absorbed by the particle detector. 
The goal of the Letter is to explore this radical scenario in detail. At the end we will see that (C) offers a possible way out of the major conceptual difficulties of quantum mechanics. 
The first part of the Letter analyzes the possibility (C) within the context of the double-slit experiment. The second, more technical part does the same in the context of the EPR experiment.    

To begin with, one needs a positive statement consistent with (C). Indeed, as stated, the possibility (C) cannot be used in a constructive way. In such a form (C) is in fact consistent with a stronger form of the Copenhagen interpretation that denies existence of the electron before it is absorbed by a particle detector. The exact opposite position will be taken here. Namely, it will be assumed that {\em the electron in the double-slit experiment exists (in some physical form) throughout the entire experiment.}

From this assumption, (C) and the topology of the space $R^{3}$ divided by the screen with the slits, one concludes that between emission and absorption the electron lives outside $R^{3}$. So the only way to reconcile the above statements is by giving up the common perception that the electron is ``attached'' to the classical space. Instead, this space itself must be a ``part'' of a physical space of more dimensions, into which the electron can escape. 

To develop this thought, recall that the electron's state in quantum mechanics is captured by the wave function. In the case when one is interested only in the electron's position, this is a complex-valued function of spatial coordinates. The evolution of the electron in the double-slit experiment is given by a path $\varphi_{t}$ with values in a Hilbert space $H$ of such functions. This path originates at a point $\varphi_{t_{1}}$, given by the wave function of the electron that is about to be emitted by the source. If the source is located at ${\bf a} \in R^{3}$, the corresponding point $\varphi_{t_{1}}$ in $H$ is ideally the Dirac delta function $\delta^{3}({\bf x}-{\bf a})$ \footnote{ See \cite{Kryukov3} for examples of Hilbert spaces containing delta functions and related mathematical considerations.}.  The end-point $\varphi_{t_{2}}$ of the path is given by the wave function of the electron absorbed by the particle detector. If the electron was detected at ${\bf b} \in R^{3}$, then $\varphi_{t_{2}}({\bf x})=\delta^{3}({\bf x}-{\bf b})$. 

Assuming $H$ contains the set $M_{3}$ of all delta functions $\delta^{3}({\bf x}-{\bf u})$ with ${\bf u} \in R^{3}$, one can identify the classical Euclidean space $R^{3}$ with the set $M_{3}$.
Indeed, there is an obvious one-to-one correspondence between $R^{3}$ and $M_{3}$ via the map $\omega:{\bf u} \longrightarrow \delta^{3}({\bf x}-{\bf u})$. This correspondence is physically meaningful: if the electron is located at a point ${\bf u} \in R^{3}$, then the electron's wave function is the eigenstate $\delta^{3}({\bf x}-{\bf u})$ of the position operator ${\widehat {\bf x}}$ and vice versa  \footnote{The phase of the delta functions in the identification is taken to be zero.}.
Moreover, for an appropriately chosen Hilbert space $H$ the map $\omega$ is an {\em isometric embedding} which means that $R^{3}$ and $M_{3}$ are identical manifolds with a metric.
As a result, the classical Euclidean space $R^{3}$ can be identified in a physically meaningful way with the submanifold $M_{3}$ of a Hilbert space $H$ of wave functions \footnote{The embedding $\omega$ can be used to {\em derive} the classical space from a Hilbert space of states \cite{Kryukov}.}.
Accordingly, the following statement will be accepted: 
\begin{description}
\item [(S)]
The classical space arena in physics is a part of a larger, Hilbert space arena. 
Physical processes with an electron on the classical space $R^{3}$ are particular cases of physical processes on the Hilbert space $H$ of the electron's states. 
More precisely, the classical space $R^{3}$ can be physically identified with the manifold $M_{3}$ of the wave functions $\delta^{3}({\bf x}-{\bf u})$, ${\bf u} \in R^{3}$ of the electron. The evolution of the electron is a path $\varphi_{t}$ in the space $H$. Whenever the electron is detected at a point ${\bf a}$ in $R^{3}$, the electron's path $\varphi_{t}$ passes through the point $\delta^{3}({\bf x}-{\bf a})$ in $M_{3}$.
\end{description}
                         
How does the statement (S) help interpret the double-slit experiment? It answers the question of what happens to the electron between its emission and absorption. Recall that the initial and the terminal points of the electron's path $\varphi_{t}$ are in the classical space $R^{3}$ (identified with the manifold $M_{3}$). However, between these points the electron is in a superposition of the delta-like  states. 
Such a superposition is not given by a delta function and therefore is {\em not} a point in the classical space. So the electron's path begins at a point in the classical space, then leaves this submanifold while staying in the space of states, passes {\em over} the screen with the slits (located in classical space) and then returns to the classical space as it is absorbed by the detector.  

A specific form of the path $\varphi_{t}$ depends on details of interaction of the electron with the source, the detector and the screen with the slits. However, any such path consists of the same basic segments: propagation from the source toward the screen with the slits, passing ``through'' the slits, propagation behind the screen toward the detector, and ``collapse'' on the detector. The first three of these segments are shown in Fig. 1.
\begin{figure}[ht]
\label{fig:1}
\epsfxsize=2.7cm
\centerline{\epsfbox{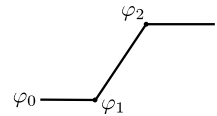}}
\caption{\small{Passing through the slits as a refraction of $\varphi_{t}$}}
\end{figure}
In the figure, the horizontal segments represent the propagation of the electron toward and away from the screen with the slits. The middle segment represents the motion of the electron ``through'' the slits when the initial electron's wave packet ``splits'' into a superposition of two wave packets. 

Whatever the actual form of the path in Fig. 1 may be, it is clearly {\em single-valued} and {\em continuous}, i.e., it is a path in the mathematical sense. In particular, for each value of the parameter $t$ there is only one point $\varphi_{t}$ in $H$. The screen with the slits simply causes a {\em refraction} of the electron's path. Notice the stunning difference between Fig. 1 and the standard picturing of the double-slit experiment shown in Fig. 2. 
\begin{figure}[ht]
\label{fig:2}
\epsfxsize=3.2cm
\centerline{\epsfbox{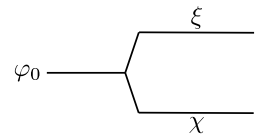}}
\caption{\small{The standard picturing of the double-slit experiment}}
\end{figure}
In the figure, $\xi, \chi$ are the wave functions of the electron passing through one of the slits with the second slit closed. They are the (normalized) components of the superposition $\psi=c_{1}\xi+c_{2}\chi$ that represents the electron behind the slits.
The splitting of the electron's path in Fig. 2 is due to attaching the path to the classical space and is responsible for the paradox associated with the experiment. Namely, 
by insisting that the electron is in the classical space $R^{3}$ one is forced to accept that the electron goes along {\em two} different paths in $R^{3}$. That is, both components $\xi$, $\chi$ must be real. This by itself is contradictory. In fact, if the same wave function is written as a superposition of eigenstates of a different observable, then, by the same logic, the new components must be real as well. Since there are many observables, the notion of reality becomes ill-defined. This is known as a ``preferred basis problem'' in quantum mechanics. 
A very similar situation arises in classical physics. Namely, when a physical vector (say, a velocity vector) is written in terms of its 
components in a certain basis, should we count the components as real? The answer is obvious: the physical vector itself is real because it is basis independent. However, relative to the vector, the components are just ``shadows'' of the real thing as they change with the change of basis, very much like shadows change when the sources of light are moved around. In quantum mechanics too, one should say that the state itself is a ``real thing'', while the components are not. 

For instance, consider the event of passing through the slits in one dimension with the $X$-axis along the screen with the slits and orthogonal to the slits. If $\delta_{x_{1}}(x)\equiv \delta(x-x_{1})$, $\delta_{x_{2}}(x)\equiv \delta(x-x_{2})$ are the (idealized) wave functions of the electron passing through the slits at $x=x_{1}$ and $x=x_{2}$ respectively, then the wave function of the electron ``passing through both slits'' is a superposition $c_{1}\delta_{x_{1}}+c_{2}\delta_{x_{2}}$ with $c_{1}, c_{2} \neq 0$. This superposition is the actual state of the electron in one dimension right behind the screen with the slits, so it is a ``real thing''. On the other hand, the components $\delta_{x_{1}}$, $\delta_{x_{2}}$ themselves,  no matter how familiar and real they seem to us, are only secondary and ``representation dependent''. So instead of having two ``real'' components one has now a single superposition. Instead of having two ``real'' electron's paths one now has a {\em single} path in the space of states. Instead of passing through both slits at once, the electron in the experiment does {\em not} pass through either of them.

The physical meaning of the superposition $\psi=c_{1}\delta_{x_{1}}+c_{2}\delta_{x_{2}}$ in the double-slit experiment is now transparent and consistent with its mathematical meaning.
Namely, the superposition is the decomposition of the actual state in a basis. Once again, it is wrong to think that the components $\delta_{x_{1}}$, $\delta_{x_{2}}$ of this decomposition are real, while the actual state $\psi$ is not. Rather, the exact opposite is true in the experiment. So, instead of {\em superposing} the ``real'' states $\delta_{x_{1}}$, $\delta_{x_{2}}$ to obtain a state that is not real,  one {\em decomposes} the actual state into the components that do not enjoy an independent existence in the experiment.  

More generally, the superposition principle is essentially similar to writing equations of the classical particle mechanics in components. However, instead of dealing with the motion of a classical particle along a path ${\bf x}_{t}$ in the Euclidean space $R^{3}$ one deals now with the motion of the electron along a path $\varphi_{t}$ in the space of states $H$. Instead of representing ${\bf x}_{t}$ by its components in an appropriate basis, one now does the same for the path $\varphi_{t}$. 
In this sense quantum mechanics becomes an extension of the classical particle mechanics onto the space of states.

One may wonder how this ``mechanical'' motion may account for the wave-like properties of the electron. The answer is simple: these properties follow from the functional nature of the points on the path $\varphi_{t}$. For instance, when two wave packets are superposed, the square of the modulus of the resulting state $\psi$ contains the interference term. So, when $\varphi_{t}$ passes through the point $\psi$, the electron behaves like a wave. 
One concludes that the ``wave-particle duality'' is completely captured by the electron's motion in the space of states. Whenever the path crosses the classical space $M_{3}$, we see it as a particle. Whenever it leaves $M_{3}$ and passes through the regions represented by less localized states in $H$, it behaves like a wave. 

The final part of the electron's evolution in the double-slit experiment is the collapse on the detector behind the slits. 
The collapse is mysterious because of its apparent discontinuity and non-locality.
In particular, how could the electron's wave function, which is in general non-vanishing over large distances in $R^{3}$, instantaneously ``shrink'' to a point supported state? Also, how could finding the electron in one place instantaneously affect results of measurements at a distant place? This is especially paradoxical if one thinks of the electron in the state $\varphi({\bf x})$ as a ``cloud'' of the density $\left|\varphi({\bf x})\right|^{2}$. Once again, the Copenhagen interpretation discards this problem by saying that no such cloud can be observed and no superluminal signaling based on the collapse can be achieved in experiments. In other words, the Copenhagen school denies collapse a physical status. However, here the electron is assumed to exist throughout the entire experiment. Consequently, the collapse in the experiment must be described both mathematically and physically. 

Consider once again the wave function $\psi=c_{1}\delta_{x_{1}}+c_{2}\delta_{x_{2}}$ of the electron right behind the slits in one dimension. Suppose that under a measurement this function collapses into $\delta_{x_{1}}$. In this case the electron is assumed to pass through the first slit. 
The paradox of collapse resides once again in thinking that both terms of $\psi$ represent a reality. That is, that the electron is in both places at once. Because of that the process of collapse seems to require an instantaneous transfer of the electron from $x_{2}$ to $x_{1}$. However, the electron in the state $\psi$ is at {\em neither} of these two places. 
Rather, it is at the point $\psi$ in $H$ which is not on the classical space submanifold of $H$.  
The collapse is {\em not} a process on the classical space. It {\em does not} collect the electron's pieces into a particle. Indeed, there are no pieces to collect!  
Rather, the electron is represented by a {\em single point} in the space of states and collapse is a motion $\varphi_{t}$ that connects that point to a point in the classical space. The fact that the electron was found at $x_{1}$ does {\em not} mean that it has passed through the first slit. Instead, provided both slits were open, the electron did not pass through either of them!

This resolves the paradox of the delayed-choice experiment, when the decision to determine 
``which slit the electron went through'' is made {\em after} the electron interacted with the 
slits. Such a delayed measurement is known to destroy the interference pattern on the photographic 
plate. The paradox is: how could the pattern disappear if the measurement 
occurred {\em after} the electron has already ``made up its mind'' and ``passed through both slits''?
The answer is now obvious: whether or not the measurement occurred, the electron in the experiment 
did {\em not} pass through the slits and the segments of the electron's path in Fig. 1 did not change. If the measurement (delayed or not) occurred, the collapse segment is added to the path. This transforms the superposition into a single concentrated packet and destroys the interference pattern on the plate behind. 

One can see that collapse is in a way opposite to the process of passing ``through'' the slits. Namely, whereas the screen with the slits ``splits'' a wave packet into a superposition of two packets, the collapse reduces the superposition into a single packet. Whereas the slits ``push'' the electron away from the classical space, the collapse returns it back to that space. 
In this respect collapse on the detector is another refraction of the electron's path in the space of states. The full path of the electron in the double-slit experiment is shown in Fig. 3.
\begin{figure}[ht]
\label{fig:3}
\epsfxsize=4cm
\centerline{\epsfbox{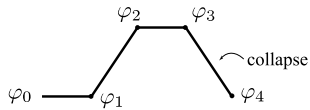}}
\caption{\small{Double-slit experiment with collapse as a path in $H$}}
\end{figure}

The  collapse segment of the path in Fig. 3 is shown to be continuous. This seems to contradict the known discontinuous, nonlocal character of collapse. For instance, how could a continuous process account for the instantaneous effect that finding the electron at $x_{1}$ has on the wave function and measurements at a possibly distant point $x_{2}$?
Recall however that collapse is happening on the space of states. 
So, instead of worrying about the distance between $x_{1}$ and $x_{2}$ in the classical space, one should worry about the distance between the points $c_{1}\delta_{x_{1}}+c_{2}\delta_{x_{2}}$ and $\delta_{x_{1}}$  {\em in the space of states}. Instead of asking about the speed of collapse in the classical space, one should ask about the speed of the evolution $\varphi_{t}$ {\em in the space of states}. 
This shift allows one to model the discontinuous, non-local process of collapse on the classical space by a continuous, local process on the space of states.

The proof of this statement is given in the second part of the Letter. It is based on the fact that
the distance between $c_{1}\delta_{x_{1}}+c_{2}\delta_{x_{2}}$ and $\delta_{x_{1}}$ can be small even if the distance between $x_{1}$ and $x_{2}$ is large. Namely, 
the correspondence $\omega$ allows one to identify the classical space with a spiral-like submanifold $M_{3}$ of an arbitrarily small sphere $S^{H}$ in the space of states $H$ (see part 2). Under the embedding, the infinite ``size'' of the Euclidean space $R^{3}$ has its counterpart in the infinite dimensionality of $S^{H}$ rather than its radius \cite{Kryukov3}. 

Take the radius of the sphere $S^{H}$ to be, say, one Planck unit of length ($\approx 1.6\cdot 10^{-35}$m). Then the distance between any two states on $S^{H}$ is at most $\pi$ Planck  units. 
Assume that collapse is the motion along a geodesic $\varphi_{t}$ between the initial and the terminal states on $S^{H}$ \cite{Kryukov2}. 
Since geodesics are continuous curves, the path $\varphi_{t}$ is {\em continuous}. Also, because the equation of geodesics is a differential equation, the metric on a small neighborhood of a point is sufficient to find the path $\varphi_{t}$ near that point. In other words, collapse is modeled by a {\em continuous local process} on the sphere of states.
Suppose now that the speed of collapse on the sphere of states is equal to the speed of light. Then the collapse from $c_{1}\delta_{x_{1}}+c_{2}\delta_{x_{2}}$ onto $\delta_{x_{1}}$ happens in less than $10^{-43}$s for {\em all} values of $x_{1}$ and $x_{2}$! 
The same process identified with a propagation from $x_{1}$ to $x_{2}$ in the classical space would require an infinite speed and would be a discontinuous action at a distance.

So, does this resolve the paradox of the double-slit experiment? Not quite. The presented analysis of the experiment was kinematical. The collapse process was assumed to be a geodesic motion on the sphere of states, but that assumption was not developed (see \cite{Kryukov2} for results in this direction). Despite of this, the provided analysis addresses the questions that normally appear in any discussion of the double-slit experiment (see for example the famous treatment by Feynman \cite{Feynman}). The analysis demonstrates that the paradigm shift from the classical space to the space of states resolves the paradox {\em in principle}.

\section{The EPR experiment}
\setcounter{equation}{0}

The Bohr-Einstein debate on the meaning of quantum theory culminated in the famous EPR paper \cite{EPR} followed by Bohr's reply \cite{Bohr}. At this time the issue of completeness of the theory was at stake. The completeness turned out to be dependent on the physical quantities designated to be {\em real} in the theory. In \cite{EPR} EPR consider a pair of non-interacting entangled particles. The state function of the pair is such that, given the position or momentum of the first particle one can predict the position or momentum of the second. Because the particles do not interact, EPR argue that both position $q$ and momentum $p$ of the second particle must be real. 
On the other hand, quantum mechanics denies that $q$ and $p$ can be simultaneously determined. Accordingly, EPR conclude that the quantum-mechanical description of reality is incomplete.

In his reply \cite{Bohr}, Bohr denies that position and momentum of a particle may be simultaneously real.
He argues that the measuring instrument itself {\em defines} the reality of {\em either} position {\em or} momentum of the particle and that the quantum-mechanical description is complete. The two points of view can be summarized as follows.
\begin{description}
\item[(A)]
{\em Both, the position and the momentum of the particle in the example are real. Quantum-mechanical description of reality is incomplete}.
\item[(B)]
{\em The position and momentum of the particle cannot be simultaneously real. The measuring device itself defines the reality of one or the other. Quantum-mechanical description of reality is complete}.
\end{description}
Bohr's position is generally adhered within the quantum community. It comes at a price of accepting that the reality of either position or momentum of the second (possibly distant) particle may be decided instantaneously by a measurement performed on the first particle. The resulting ``spooky action at a distance'' or quantum non-locality has never been acknowledged by Einstein and remains a mystery of the theory. 

In this part of the Letter a definition of reality that is capable of resolving the paradox of quantum non-locality will be proposed. This definition is consistent with Bohr's conclusion that either $q$ or $p$ but not both are real in the EPR example. At the same time it disagrees with the Bohr's  positivist's statement that the measuring device {\em defines} the reality. In this it is in line with the EPR realist's attitude and the statement of incompleteness of quantum description. 

Recall that a single spinless particle found at a point ${\bf u} \in R^{3}$ is described in quantum mechanics by the eigenfunction $\delta^{3}_{{\bf u}}({\bf x}) \equiv \delta^{3}({\bf x}-{\bf u})$ of the position operator ${\widehat {\bf x}}$. Moreover, there is an obvious one-to-one correspondence $\omega$ between $R^{3}$ and the set $M_{3}$ of all delta functions $\delta^{3}_{{\bf u}}$ via $\omega: {\bf u} \longrightarrow \delta^{3}_{{\bf u}}$. Hence, $\omega$ maps points in the classical space to states of the particle located at these points. As shown below, for an appropriate Hilbert space $H$ the map $\omega$ is an {\em isometric embedding}, which means that $R^{3}$ and $M_{3}$ are identical manifolds with a metric. In other words, the classical Euclidean space $R^{3}$ can be identified in a physically meaningful way with a submanifold of the Hilbert space of states $H$.

Consider now a pair of distinguishable particles such that the first particle is located at a point ${\bf u}$ and the second at a point ${\bf v}$ in $R^{3}$. In classical mechanics such a pair is described by a single point $({\bf u}, {\bf v})$ in the configuration space $R^{6}=R^{3}\times R^{3}$. In quantum mechanics the pair is described by the point 
$\delta^{3}_{{\bf u}}\otimes \delta^{3}_{{\bf v}}({\bf x_{1}}, {\bf x_{2}})\equiv \delta^{3}_{{\bf u}}({\bf x_{1}})\delta^{3}_{{\bf v}}({\bf x_{2}})$ in the tensor product space $H\otimes H$. Here $H$ is the space of states of one of the particles, which is assumed to be the same for both particles. Given the right $H$, the one-to-one map $\omega \otimes \omega: ({\bf u}, {\bf v}) \longrightarrow \delta^{3}_{{\bf u}}\otimes \delta^{3}_{{\bf v}}$ identifies the configuration space $R^{6}$ with the six dimensional submanifold $M_{6}$ of $H \otimes H$ consisting of the state functions $\delta^{3}_{{\bf u}}({\bf x_{1}})\delta^{3}_{{\bf v}}({\bf x_{2}})$. As before, the map $\omega \otimes \omega$ is physically meaningful as it identifies each pair of points in the classical space $R^{3}$ with the state of the pair of particles located at these points. 

Recall that a single particle with momentum ${\bf p}$ is given in quantum mechanics by the eigenstate $e^{i{\bf p}{\bf x}}$ of the momentum operator ${\widehat {\bf p}}$. Consider the subset ${\widetilde M_{3}}$ of the space of states $H$ consisting of the functions $e^{i{\bf p}{\bf x}}$ with ${\bf p}\in R^{3}$. Once again, for an appropriate realization of the space of states $H$ the map $\rho: {\bf p} \longrightarrow e^{i{\bf p}{\bf x}}$ is an isometric embedding of the classical momentum space $R^{3}$ into the space of states. One can similarly consider the space $R^{6}$ of pairs $({\bf p}, {\bf q})$ of momenta of two particles. The map $\rho \otimes \rho: ({\bf p}, {\bf q}) \longrightarrow e^{i{\bf p}{\bf x_{1}}}e^{i{\bf q}{\bf x_{2}}}$ identifies $R^{6}$ with the submanifold ${\widetilde M_{6}}$ of $H\otimes H$ consisting of the state functions $e^{i{\bf p}{\bf x_{1}}}e^{i{\bf q}{\bf x_{2}}}$. The embeddings $\rho$ and $\rho \otimes \rho$ are both physically meaningful as they identify the momentum of each particle with the corresponding state. 
Note that the classical phase space of the pair cannot be embedded in such a way into the space of states. This is because there is no state in $H$ for which both position and momentum of a particle are defined. Geometrically speaking, the intersections $M_{3}\cap{\widetilde M_{3}}$ and $M_{6}\cap{\widetilde M_{6}}$ are empty.

The maps $\omega$, $\rho$, $\omega \otimes \omega$, $\rho \otimes \rho$ allows one to identify the physical quantities of position and momentum of a particle or a pair of particles with the variable $\varphi$ taking values in one of the manifolds $M_{3}$, ${\widetilde M_{3}}$, $M_{6}$, ${\widetilde M_{6}}$. Suppose that the state function $\varphi$ itself is the most appropriate way of describing the reality. Note in particular that: {\em (1) the state function yields the most complete description of quantum system and its evolution; (2) the state function is the ``smallest'' object that provides such a complete description in a sense that it contains only the experimentally verifiable information; (3) in special cases the knowledge of state function is equivalent to the knowledge of precise position or momentum of particles in the system}.
In fact, (1) is known to be true in quantum mechanics, (3) was already discussed. As for (2), 
note that in principle, given sufficiently many copies of the system, one can experimentally determine the modulus and the phase (up to a constant initial phase) of the state function as precisely and one wishes. So, (2) is an accurate statement as well. 
The following alternative to the above positions (A), (B) is then proposed:
\begin{description}
\item[(C)]
{\em Physical reality of a pair of particles is most appropriately described by the state variable $\varphi$ of the pair. The evolution of the pair in time is a path $\varphi_{t}$ in the space of states. The variable $\varphi$ generalizes the classical positions and momenta of the particles and reduces to those in special cases. Neither positions nor momenta of the particles are generally defined. The positions are defined if and only if $\varphi$ takes values in the submanifold $M_{6}$ of the space of states $H \otimes H$ of the pair. The momenta are defined if and only if $\varphi$ takes values in the submanifold ${\widetilde M_{6}}$ of $H \otimes H$. Because the intersection $M_{6}\cap{\widetilde M_{6}}$ is empty, the positions and momenta cannot be simultaneously defined. The process of measurement does not create a reality: the state exists before and after the measurement. Rather, similarly to any interaction, a measurement simply moves the state. In particular, a measuring device that measures positions of the particles brings the initial state $\varphi$ to a point of $M_{6}$. Similarly, a device that measures momenta of the particles forces the state onto ${\widetilde M_{6}}$}.
\end{description}  

How does statement (C) help understand the EPR experiment? EPR consider a pair of particles in one dimension in an entangled state given by the state function $\varphi(x_{1}, x_{2})=\frac{1}{2\pi\hbar}\int^{\infty}_{-\infty}e^{\frac{i}{\hbar}(x_{1}-x_{2}+x_{0})p}dp$, where $x_{0}$ is a constant. From the form of $\varphi$ one can see that whenever the position of the first particle is known to be $u$, the position of the second must be $x_{0}+u$. Similarly, whenever the momentum of the first particle is $p$, the momentum of the second must be $-p$ (see \cite{EPR}). Let $H$ be the Hilbert space of states of each particle so that $\varphi$ is in $H\otimes H$. Note that neither the position nor the momentum of the particles in this state is defined. In the geometric terms that means, once again, that $\varphi$ does not belong to the submanifolds $M_{6}$ or ${\widetilde M_{6}}$ of $H\otimes H$. After the measurement of position of the first particle, the state $\varphi$ moves to a point of $M_{6}$. Similarly, the measurement of momentum of the first particle brings the state to ${\widetilde M_{6}}$. So, the system moves from the state in which neither position nor momentum of the particles is real to a state in which either position or momentum (but not both) of the particles is real. This is of course consistent with the Bohr's interpretation. However, for Bohr the act of measurement defines the reality. In particular, no physical description of collapse is possible.  
Here on the other hand, the reality is defined by the state. 
Because the state exists before and after the measurement, it becomes possible to analyze the collapse both mathematically and physically. In particular, it becomes possible to address the paradoxes of the EPR experiment. 

To see how this can be done, let's express the state function $\varphi$ of the EPR-pair in the form
\begin{equation}
\label{delta}
\varphi(x_{1},x_{2})=\int \delta_{u}(x_{1})\delta_{x_{0}+u}(x_{2})du.
\end{equation}
In this form the state $\varphi$ is a superposition of all states $\delta_{u}(x_{1})\delta_{x_{0}+u}(x_{2})$ that correspond to the first particle being at a point $u$ and the second at the point $x_{0}+u$.
In discussing the EPR experiment one usually makes a tacit assumption that the two particles are always located in the classical space. The fact that the superposition in (\ref{delta}) contains various terms $\delta_{u}(x_{1})\delta_{x_{0}+u}(x_{2})$ signifies then that the particles are located at {\em all} pairs of points $(u, x_{0}+u)$ at once. That is, the particles must somehow ``split'' between these points.
This thinking leads one to the conclusion that measuring position of the first particle we somehow ``collect'' the particle into a single point and pass this information through the classical space to the second particle so that it could also ``assemble'' at a predetermined point. This is certainly paradoxical! So, {\em how could a measurement of position $x_{1}$ of the first particle instantaneously fix the position $x_{2}$ of the second, possibly distant particle?}

According to (C), the reality is not given by the components $\delta_{u}(x_{1})\delta_{x_{0}+u}(x_{2})$ of $\varphi$, but rather by the state $\varphi$ itself. So the pair does not ``split'' between various points in the classical space but is given instead by a {\em single} point $\varphi$ in the space of states $H \otimes H$, away from the submanifold $M_{6}$. The fact that $\varphi$ is not a product $\xi(x_{1})\chi(x_{2})$ of two functions signifies that the reality before the collapse cannot be described in terms of individual particles. Instead, the state function $\varphi$ of the {\em pair} provides the only adequate representation of reality. 
Furthermore, to measure position of the ``first particle'' is to bring the pair represented by $\varphi$ to $M_{6}$. Indeed, by definition of $\varphi$, whenever the first particle is at the point $x_{1}=a$, the second particle is at $x_{2}=x_{0}+a$. Consequently, the state function of the pair after the measurement is $\delta_{a}(x_{1})\delta_{x_{0}+a}(x_{2})$, which is a point in $M_{6}$. 
So instead of collecting pieces of the particles, spread over the classical space, the process of collapse moves the pair from the point $\varphi$ onto the manifold $M_{6}$. Thus, the process of collapse is a path $\varphi_{t}$ in the space of states that connects the point $\varphi$ in $H\otimes H$ to the point $\delta_{a}\otimes \delta_{x_{0}+a}$ in $M_{6}$. 

How could the collapse happen instantaneously even when the particles are far apart?
Because the process of collapse is happening on the space of states $H\otimes H$ and not on the classical space, 
the spatial distance between the particles is irrelevant. What matters now is the distance between the states $\varphi$ and $\delta_{u}\otimes \delta_{x_{0}+u}$ and the speed of evolution $\varphi_{t}$ {\em in the space of states}. 
Importantly, the distance between the states may be small even when the distance between the particles is known to be large. Indeed, as shown below, for an appropriate space of states $H$ the map $\omega \otimes \omega$ identifies the classical configuration space $R^{6}$ with a submanifold of an arbitrarily small sphere $S^{H\otimes H}$ in $H\otimes H$. Accordingly, the distance between any two states may be arbitrarily small.
In this case a finite speed of the evolution $\varphi_{t}$ on the space of states may be perceived as an instantaneous process on the classical space. Namely, it is claimed that 
\begin{description}
\item[(S)]
An apparently discontinuous, nonlocal process of collapse on the classical space can be modeled by a continuous, local process on the space of states.  
\end{description}

Before proving (S), let's address yet another mystery of the EPR pair: 
{\em How could the reality of either position or momentum of the second particle be instantaneously determined by the observer's decision to measure position or momentum of the first particle?}
Once again, the key to resolving this mystery is to observe that before the measurement, the pair of the particles is {\em not} located in the classical configuration space $M_{6}$ or the classical momentum space ${\widetilde M_{6}}$. In particular, the particles are {\em not} spread over all possible positions or momenta. (This by itself would be contradictory. Indeed, should the particles be spread over possible positions or possible momenta? If that depends on a measurement, then how would a particular spreading be created?)
Under the position measurement ``on the first particle'', the entire pair moves along a path $\varphi_{t}$ from the point $\varphi$ to a point $\delta_{a}(x_{1})\delta_{x_{0}+a}(x_{2})$ on the submanifold $M_{6}$ of $H \otimes H$. Likewise, the momentum measurement on the first particle brings the pair along a different path ${\widetilde \varphi_{t}}$ from $\varphi$ to a point $e^{iqx_{1}}e^{-iqx_{2}}$ on the submanifold ${\widetilde M_{6}}$ of $H\otimes H$. So a particular measuring device (either a man-made instrument or a natural phenomenon) moves the pair to either $M_{6}$ or ${\widetilde M_{6}}$. The discontinuous, nonlocal nature of the collapse can be now explained via statement (S).

To prove (S) it suffices to provide a specific model satisfying the statement. 
For this, consider the Hilbert space obtained by completing the space $L_{2}(R^{3})$ of  complex-valued square-integrable functions on $R^{3}$ in the norm defined by the inner product
\begin{equation}
\label{inner}
(\varphi, \psi)_{H}=\int e^{-\frac{1}{2}({\bf x}-{\bf y})^{2}}\varphi({\bf x}){\overline \psi({\bf y})} d^{3}{\bf x}d^{3}{\bf y}.
\end{equation}
By plugging in $\varphi=\psi=\delta^{3}_{{\bf a}}$ one concludes that $H$ contains the delta functions and the norm of any delta-function is one. So $\omega$ identifies $R^{3}$ with a submanifold $M_{3}$ of a unit sphere $S^{H}$ in $H$.
Note that for finitely many different vectors ${\bf a_{k}}$ the functions $\delta^{3}_{{\bf a_{k}}}$ are linearly independent. Also, no element of $H$ is orthogonal to all delta functions. It follows that the manifold $M_{3}$ ``spirals through'' all available dimensions forming a {\em complete set} in $H$ (see Fig. 4). 
\begin{figure}[ht]
\label{fig:11}
\epsfxsize=4.5cm
\centerline{\epsfbox{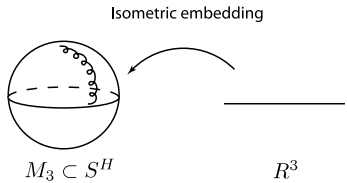}}
\caption{\small{$R^{3}$ as a submanifold of the sphere $S^{H}$}}
\end{figure}
The induced metric on $M_{3}$ is given by the components $g_{ik}=\left.\frac{\partial^{2}k(x,y)}{\partial x^{i} \partial y^{k}}\right |_{{\bf x}={\bf y}={\bf a}}$, where $k({\bf x},{\bf y})=e^{-\frac{1}{2}({\bf x}-{\bf y})^{2}}$ is the kernel of the metric (\ref{inner}). Differentiation yields the ordinary Euclidean metric, so that $M_{3}$ is identical (i.e., {\em isometric}) to the Euclidean space $R^{3}$ (see \cite{Kryukov3}).

Because the classical space $R^{3}$ is isometrically embedded into $H$, the distances on $S^{H}$ can be measured in the ordinary units of length. To make the distance between any two states on $S^{H}$ small, the unit sphere itself must be small. Accordingly, the unit of length must be small. For instance, in the Planck system of units the radius of the unit sphere $S^{H}$ is one Planck length ($\approx 1.6\cdot 10^{-35}$m). In this case the distance between any two states on the sphere (which is equal to the angle $\theta$ between the corresponding vectors in $H$) does not exceed $\pi$ Planck lengths. For example, the distance between $\delta^{3}_{{\bf a}}$ and $\delta^{3}_{{\bf b}}$ increases monotonically with $\left\|{\bf a}-{\bf b}\right\|_{R^{3}}$ and tends to $\pi/2$ Planck lengths as  $\left\|{\bf a}-{\bf b}\right\|_{R^{3}}$ tends to infinity.  Of course, when this distance is measured along the classical space ``spiral'' $M_{3}$ (rather than the great circle connecting the states), it takes arbitrarily large values, equal to the norm $\left\|{\bf a}-{\bf b}\right\|_{R^{3}}$.

Note that when the Planck system of units is used, the kernel $k({\bf x},{\bf y})=e^{-\frac{1}{2}({\bf x}-{\bf y})^{2}}$ of the metric (\ref{inner}) is an extremely sharp, practically point-supported function. Indeed, $k({\bf x},{\bf y})$ falls off to almost zero within the first few Planck lengths of $\left\|{\bf x}-{\bf y}\right\|_{R^{3}}$. That means that for the usual in applications functions, $k({\bf x},{\bf y})$ behaves like the delta function $\delta^{3}({\bf x}-{\bf y})$. By replacing the kernel $k({\bf x},{\bf y})$ in (\ref{inner}) with $\delta^{3}({\bf x}-{\bf y})$ one obtains the ordinary $L_{2}$-inner product:  $\int \delta^{3}({\bf x}-{\bf y})\varphi({\bf x}){\overline \psi({\bf y})}d^{3}{\bf x} d^{3}{\bf y}=\int \varphi({\bf x}){\overline \psi({\bf x})}d^{3}{\bf x}$. It follows that the $H$-norms of the usual square-integrable functions are extremely close to their $L_{2}$-norms. That verifies  that the Hilbert space $H$ is {\em physical}, i.e., it can be consistently used in quantum mechanics in place of the ordinary space $L_{2}(R^{3})$.

Consider first the collapse of a single particle state under a measurement of the particle's position.
Assume that collapse is the motion along a geodesic $\varphi_{t}$ connecting the initial and the terminal states on $S^{H}$ \cite{Kryukov2}. 
Because geodesics are continuous curves, the path $\varphi_{t}$ is {\em continuous}. Also, because the equation of geodesics is a differential equation, the metric on a small neighborhood of a point is sufficient to find the path $\varphi_{t}$ near that point. 
In other words, the collapse is in this case a {\em continuous local process} on the sphere of states.
Suppose now that the speed of collapse on the sphere of states is equal to the speed of light. Recall that the distance between any two states on $S^{H}$ does not exceed $\pi$ Planck lengths. It follows that the collapse of an {\em arbitrary} initial state onto an {\em arbitrary} terminal state happens in less than $10^{-43}$s. 
For instance, the collapse from the superposition $c_{1}\delta^{3}_{{\bf x_{1}}}+c_{2}\delta^{3}_{{\bf x_{2}}}$ of two position eigenstates of a particle onto the state $\delta^{3}_{{\bf x_{1}}}$ of the particle found at the point ${\bf x_{1}}$ happens in less than this time interval for {\em all} values of ${\bf x_{1}}$ and ${\bf x_{2}}$ at once! 
On the other hand, if the process of collapse is supposed to propagate in the classical space from $x_{1}$ to $x_{2}$ at a constant speed, then that speed must be infinite.
Even with the limitation that the distance between $x_{1}$ and $x_{2}$ does not exceed the size of the universe ($\approx 10^{27}$m), the above time interval would still require the collapse to have a ridiculous speed of $\approx 10^{70}$m/s! By all standards the resulting process 
is a discontinuous action at a distance. 

In the case of a position measurement on a pair of particles consider the tensor product $H\otimes H$ with the above space $H$.
The norm of the state $\delta^{3}_{{\bf a}}\otimes \delta^{3}_{{\bf b}}$ in $H\otimes H$ is the product of the $H$-norms of each delta-function, so it is equal to one. Accordingly, the set $M_{6}$ forms a submanifold of the unit sphere $S^{H\otimes H}$ in $H\otimes H$. As before, this sphere can be made small by using the Planck scale, in which case the previous consideration applies. 
To include the collapse due to a measurement of momentum of a particle or a pair of particles, one must change $H$ so that to include the eigenstates of the momentum operator. In particular, by changing the kernel of the metric (\ref{inner}) to $e^{-\alpha{\bf x}^{2}}e^{-({\bf x}-{\bf y})^{2}}e^{-\alpha{\bf y}^{2}}$, where $\alpha >0$, one obtains a possible such space. At the same time, for a sufficiently small coefficient $\alpha$ other earlier discussed properties remain valid. In particular, the new Hilbert space ${\widetilde H}$ remains physical and the metric induced on the submanifolds $M_{3}$ and ${\widetilde M_{3}}$ is arbitrarily close to the Euclidean metric.
The spaces ${\widetilde H}$ and ${\widetilde H}\otimes {\widetilde H}$ are appropriate for modeling collapse processes involving position and momentum measurements on a single particle or a pair of particles. In all these cases the previous model applies making collapse a continuous local process on the sphere of states that looks like an instantaneous process on the classical space. This completes the proof of statement (S).

It is generally accepted that quantum mechanical description of reality is more meager than the classical description. For instance, position and momentum of a particle in quantum mechanics do not have a simultaneous meaning. However, 
if reality is to be described by the state function of the system, the situation is reversed. For example, a complete classical description of a system of $N$ particles at a given time requires $6N$ numbers (positions and momenta of the particles). On the other hand, to identify the state of a single particle at a given time one needs in general infinitely many numbers (components of the state function in a basis). So the state gives a much richer (although different) information about the system than the classical mechanical physical quantities. 

One may wonder how could the state function description of {\em reality} be richer if the outcomes of our experiments are specific values of the physical quantities? The answer is simple: the state function contains information about {\em all} outcomes of the experiments on the system at once. 
It follows that quantum mechanics contains more information about reality than it normally gets credit for. Because the state is available to experimental determination, one should not insist that reality can only be associated with a specific outcome of a measurement. 
Rather, {\em all} possible outcomes of measurements on copies of a system identify a {\em single} reality of the system before measurement. Namely, these outcomes are {\em projections} of the reality  that identify the reality itself (i.e., the state or the {\em position} of the system in the space of states). 

The new definition of reality 
opens a way of investigating what happens to quantum system before, during, and after it has been measured.  
The fact that this can be done suggests that the current quantum mechanics is indeed incomplete. This incompleteness is not due to the lack of classicality in the quantum description. On the contrary, it originates in the lack of a consistent eradication of classicality from the basic tenets of quantum theory. 
By identifying the outcomes of measurements with the special cases of the reality associated with the state, one obtains a tool for embedding the classical into the quantum. 
The very first steps in this direction demonstrate that by properly completing the theory one can successfully resolve the mysteries of quantum mechanics and provide a much richer description of reality than the classical physics could ever hope for.


\end{document}